\documentclass[namedreferences]{solarphysics}
\usepackage[optionalrh,solaromanenum]{spr-sola-addons} 
\makeatletter
\let\subfigure\@undefined
\usepackage{graphicx}        
\usepackage{amssymb}         
\usepackage{color}           
\usepackage{url}             

\usepackage[a4paper]{geometry}
\usepackage{hyperref}
\hypersetup{colorlinks=true,linkcolor=blue,urlcolor=blue,citecolor=blue,breaklinks=true}

\usepackage{subfig}

\DeclareCaptionFormat{subfig}{\textbf{\figurename~#1#2}#3} 
\DeclareCaptionSubType*{figure}
\begin{document}
\begin{article}
\begin{opening}


\title{Using an Ellipsoid Model to Track and Predict the Evolution and Propagation of Coronal Mass Ejections} 

\author{S.~\surname{Schreiner}$^{1*}$\sep
        C.~\surname{Cattell}$^{1}$\sep
        K.~\surname{Kersten}$^{1}$\sep
	A.~\surname{Hupach}$^{1}$
       }

\runningauthor{Schreiner-S et al.}
\runningtitle{Using an Ellipsoid Model to Track and Predict the Evolution and Propagation of Coronal Mass Ejections}

   \institute{$^{1}$ School of Physics and Astronomy, University of Minnesota
                     email: \url{schr0910@umn.edu} \\ 
              $^{*}$ Now at: Department of Aerospace Engineering and Mechanics, University of Minnesota\\
             }

\begin{abstract}
We present a method for tracking and predicting the propagation and evolution
 of coronal mass ejections (CMEs) using the imagers on the STEREO and SOHO
 satellites. By empirically modeling the material between the inner core and 
leading edge of a CME as an expanding, outward propagating ellipsoid, we
 track its evolution in three-dimensional space. Though more complex empirical
 CME models have been developed, we examine the accuracy of this relatively
 simple geometric model, which incorporates relatively few physical
 assumptions, including i) a constant propagation angle and ii) an azimuthally
 symmetric structure. Testing our ellipsoid model developed herein on three
 separate CMEs, we find that it is an effective tool for predicting the
 arrival of density enhancements and the duration of each event near 1 AU.
 For each CME studied, the trends in the trajectory, as well as the radial 
and transverse expansion are studied from 0 to $\approx$.3 AU to create 
predictions at 1 AU with an average accuracy of 2.9 hours.

\end{abstract}
\end{opening}

\section{Introduction}
     \label{S-1} 
Coronal mass ejections (CMEs) are powerful, large-scale structures in the 
solar wind that can be hazardous to technology within and beyond Earth’s
 magnetosphere. CMEs are defined as ``…an observable change in coronal
 structure that (1) occurs on a time scale of a few minutes and several hours 
and (2) involves the appearance and outward motion of a new, discrete, bright,
 white-light feature in the coronagraph field of
 view"\cite{Hundhausen84,Schwenn96}.Though coronal mass ejections have
 been studied for decades, the physics responsible for their structure,
 evolution, and propagation is not entirely understood.
\par
This paper presents two methods that could prove useful in studying CMEs.
 First, we introduce a modification of a common technique for measuring the 
location of CME structures. By analyzing plots of intensity vs. position along
 an adjustable cut line through CME images, we find that we are able to
 effectively identify and measure the location of various CME structures.
 The benefit of this modification is twofold. It allows the user to (1)
 accurately calculate the uncertainty on each measurement and (2) track
 structures that do not propagate along constant position angles. 
\par
\begin{figure}[t]
	\centering
		\begin{minipage}[b]{0.9\textwidth}
			\begin{center}
				\includegraphics[width=\textwidth]{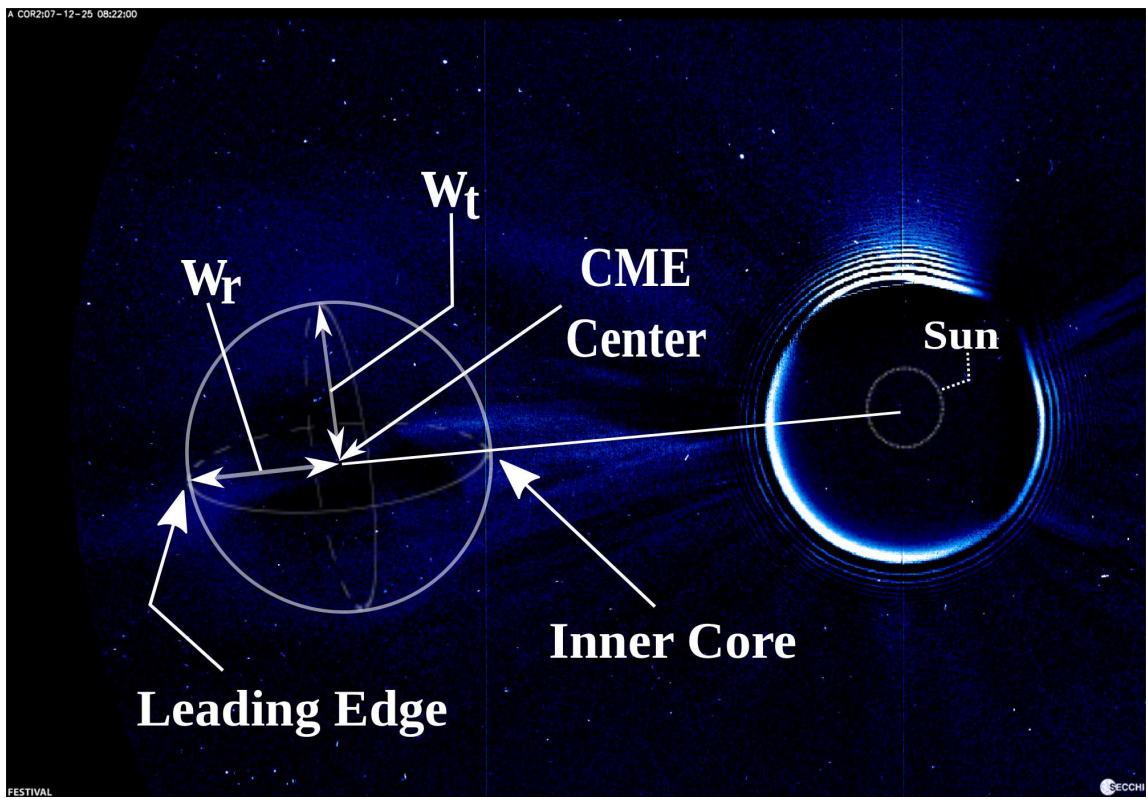}
			\end{center}
			\caption{A depiction of our ellipsoid model fit over an image from the second coronograph on STEREO A, taken on 25 December 2007. The two measureable halfwidths, the radial half-width ($w_r$), along the Sun-CME center line, and transverse halfwidth ($w_t$), perpendicular to the radial width in the image, are also show. A background subtraction and calibration have been applied.}
			\label{fig:Fig1}
		\end{minipage}
\end{figure}

Second, we conduct a preliminary assessment of the effectiveness of a
 relatively simple, empirical CME model which describes the material between
 the leading edge and inner core of a CME (herein referred to as the ``forward 
structure") as an expanding, outward propagating ellipsoid. The surface of
 this three-dimensional (3D) ellipsoid is used to approximate the outer
 density structure and inner core of CMEs, as shown in Figure~\ref{fig:Fig1}. Our 
ellipsoid model, developed herein, combines imager data from two satellites
 to generate \emph{in situ} predictions and a third satellite to provide \emph{in situ} 
observations of the CME to assess the accuracy of our predictions. 
\par
The theory of Thomson scattering predicts that the brightness seen in CME 
images is due to electron density enhancements scattering solar radiation in
 the solar corona \cite{Vourlidas06}. Therefore, tracking bright
 features that continuously appear over a large range of elongations should
 provide a reliable way to predict density enhancements in the \emph{in situ} data.
 As shown in Figure~\ref{fig:Fig1}, CMEs typically display a three-part structure:
 a bright leading edge, followed by a dark cavity, followed by a bright inner
 core \cite{Illing86}. By constructing the ellipsoid such that it
 passes through the inner core and leading edge and also spans the transverse
 width of the CME, we find that we are able to describe the leading edge and
 inner core geometry and the density enhancements associated with these
 features \cite{Vourlidas06,Lugaz08}. Figure~\ref{fig:Fig1} depicts our ellipsoid
 model fit over an image from the second coronograph on STEREO A, taken on 25
 December 2007. A background subtraction and calibration have been applied,
 which will be discussed later in the paper. The half-width of the CME in the
 radial direction, $w_{r}$, (along the Sun-CME center line) and
 transverse direction, $w_{t}$, (perpendicular to the Sun-CME center
 line) are indicated. Additionally, the bright leading edge and inner core of 
the CME and the approximate outline of the Sun are overlaid.
\par
Using images of a CME to predict the time of arrival of \emph{in situ} signatures at
 a location of interest has been the focus of much recent research
 \cite[and references therein]{Davies09,Thernisien09,Wood09}. 
\inlinecite{Wood09} modeled a CME as an inner flux-rope structure and an outer
 density structure created by the flux rope piling up solar wind material. In
 contrast, our ellipsoid model describes the outermost density structure of
 the CME, from the inner core to the leading edge, rather than the geometry of the inner flux rope. 
\par
\inlinecite{Savani09} attempted to characterize the radial and transverse 
expansion of CMEs using a circular fit in each image. Our model adds an
 additional degree of freedom by independently measuring both the radial and
 transverse width of the CME. \inlinecite{Wood11} found that the flux rope
 contained within many CMEs can be described with an elliptical cross section. 
As a first-order approximation, a flux rope with an elliptical cross section
 could be expected to drive an elliptically-curved density structure ahead of 
it.
\par
To translate the elliptical cross section from the images to a 3D model, we
 create an ellipse of revolution around the radial axis (Figure~\ref{fig:Fig1}) in the same
 way \inlinecite{Wood11} used an azimuthally symmetric outer density
 structure. This assumption is examined further in the conclusion of this
 paper.  Another study modeled the entirety of the CME structure from the Sun
 to the leading edge as a ``hollow croissant" \cite{Thernisien09}.
 Instead, our model is not anchored to the Sun but allows the forward
 structure of the CME to be modeled as an ellipsoid anchored to the leading
 edge and inner core (similar to the second model presented in
 \inlinecite{Lugaz10}). This means that our ellipsoid model describes the
 forward density structure rather than the magnetic flux-rope structure and
 therefore will not accurately model the conditions sunward of the inner 
core. By focusing on the forward structure of the CME, our model is designed 
to capture the three most prominent features in CME images: the leading 
edge, the dark cavity, and the inner core. 
\par
Our model provides a technique for tracking CMEs and predicting the arrival 
of \emph{in situ} density signatures with an average accuracy of 2.9 hours, based 
on a study of three different events. Though our study is not statistical
 in nature, these results are quite promising and corroborate the potential 
role of empirical geometric models.
\par
The paper is organized as follows. In section~\ref{S-2}, we introduce the data
 sets utilized and the methodology for creating our ellipsoid model from the
 imager data. Specific data and predictions for three separate CMEs are
 presented in section~\ref{S-3}. Comparisons to \emph{in situ} data are described in 
section~\ref{S-4}. Discussion and conclusions are given in section~\ref{S-5}.

\section{Instruments and Data Analysis Methodology} 
      \label{S-2} 
Our technique used the SECCHI imagers on the STEREO satellites \cite{Howard08}
  ahead of (ST-A) and behind (ST-B) the Earth in its orbit and the LASCO
 imagers on the SOHO satellite \cite{Brueckner95,Domingo95} at the L1 Lagrange
 point between the Earth and the Sun. Using images from these instruments, 
we tracked three CMEs out to $\approx$.3 AU and then used that data to predict the
 propagation and evolution of each CME out to 1 AU. \emph{In situ} data from the
 IMPACT magnetometer \cite{Acuna08} and PLASTIC \cite{Galvin08} instruments
 on STEREO and the MFI \cite{Lepping95} and 3DP \cite{Lin95} instruments on
 the Wind spacecraft was used to assess the accuracy of our predicted arrival
 times.
\par
Each of the three CMEs we studied was chosen because it transited STEREO A,
 STEREO B, or Wind. To make our selections, we compiled data from the STEREO
 and Wind \emph{in situ} event lists \cite{Jian10a,Jian10b}, thus ensuring that
 predictions for each chosen event could be tested with the \emph{in situ} data. 
This selection criteria allowed us to track each CME with sets of images 
from either SOHO and/or STEREO and verify the predicted arrival times using
 STEREO or Wind. It was vital to have a set of images from two different 
vantage points in order to accurately determine the propagation direction 
of the CME in 3D space. With only one set of images, one has to make various
 approximations that greatly limit the accuracy of any predictions 
\cite{Lugaz09}.  The CMEs we selected to study crossed 1 AU on
 30 December 2007 (herein referred to as ``event 1"), 29 April 2008 (``event 2"),
 and 15 December 2008 (``event 3").
\par
\begin{figure}[ht]
\centering
	\subfloat[An image of a CME from STEREO A taken on 25 December 2007 (after background subtraction and calibration). The elongation of the leading edge and inner core ($\epsilon_{LE}$, $\epsilon_{IC}$), the position angle ($\theta$), the propagation angle  ($\phi$), and the distance from the Sun to the observing satellite ($d_0$) are indicated. The modified radial, tangential, normal (RTN) coordinate system used to track the CME is also shown, centered at the Sun.]{
	\includegraphics[width=0.47\textwidth]{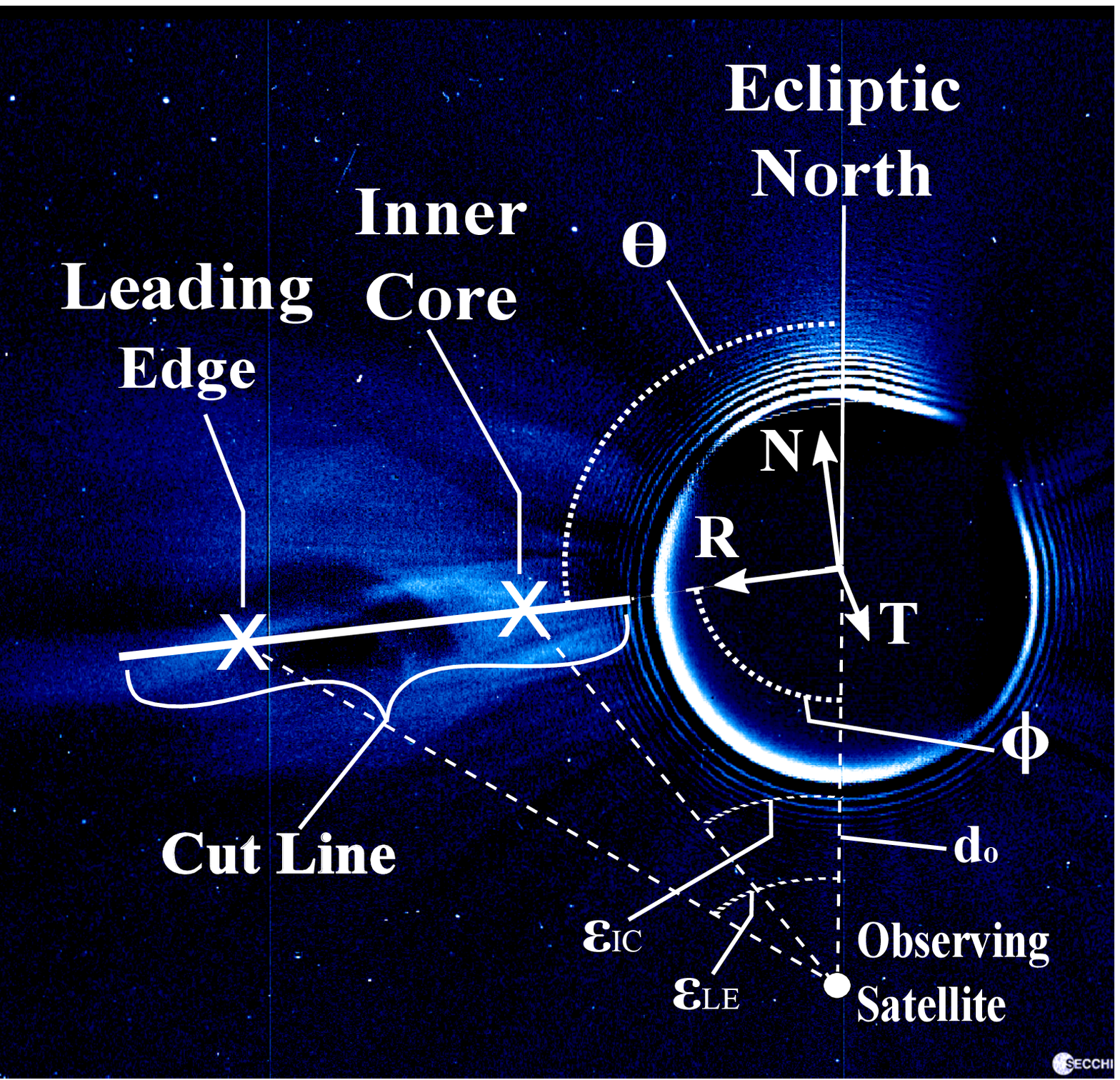}
	\label{fig:Fig2a}
}
\hfill
	\subfloat[An intensity profile (plot of longitude vs. intensity) along the white cut line in Figure 2a, generated by FESTIVAL. The locations of the leading edge and inner core are indicated with dashed lines. The spike in the data near the longitude of -2.14 degrees is due to the cut line crossing over a bright start in the image. The elongation measurements are in degrees and the intensity measurements are in units of solar disk intensity (linear scale).]{
	\includegraphics[width=0.47\textwidth]{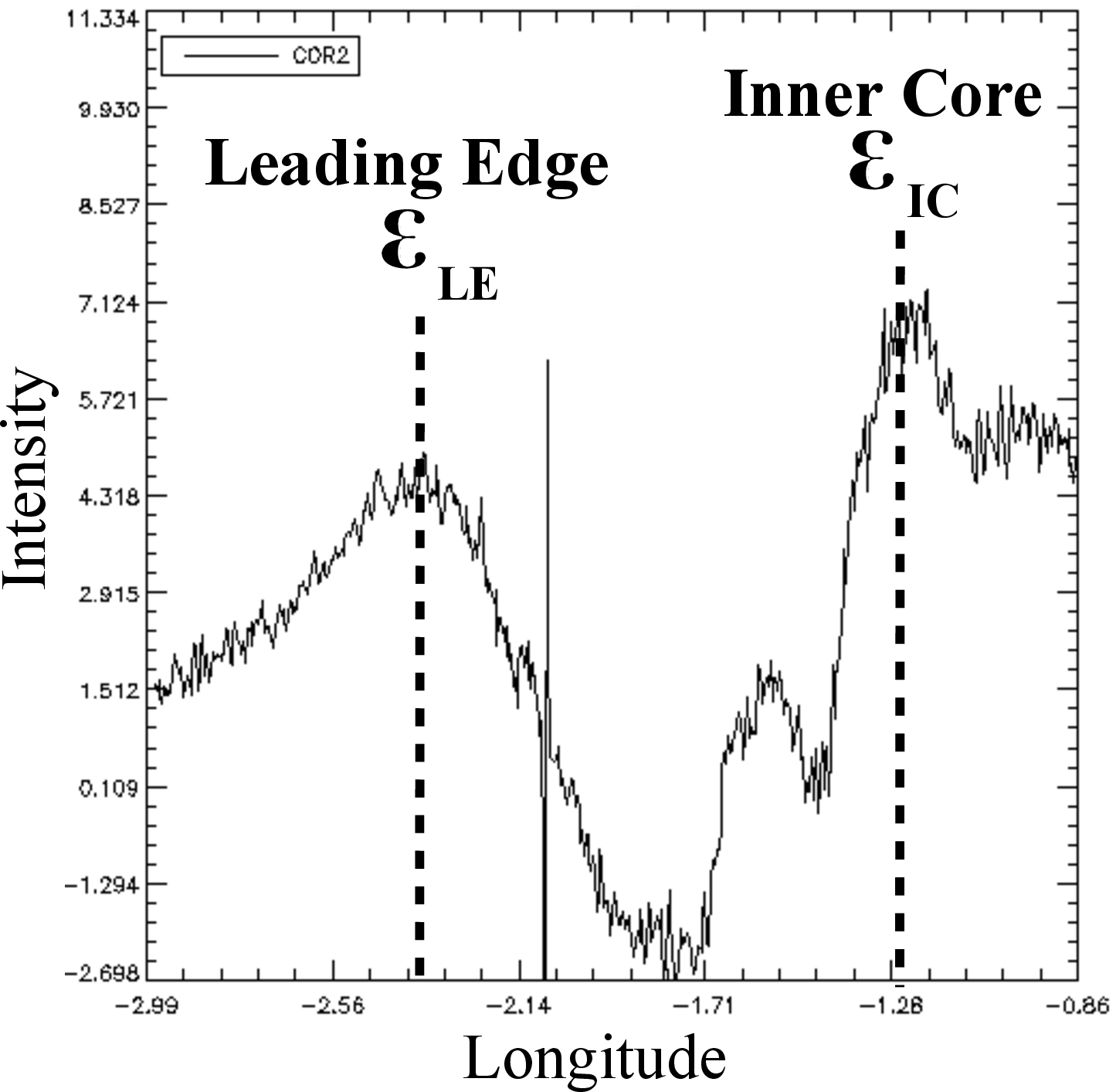}
	\label{fig:Fig2b}
}
\caption{A depiction of the characteristic variables of CME propagation (left) and a demonstration of an intensity profile (right) used to measure the elongation of the leading edge and inner core.}
\label{fig:Fig2}
\end{figure}
\begin{figure}[p]
	\centering
		\begin{minipage}[b]{0.83\textwidth}
			\begin{center}
				\includegraphics[width=\textwidth]{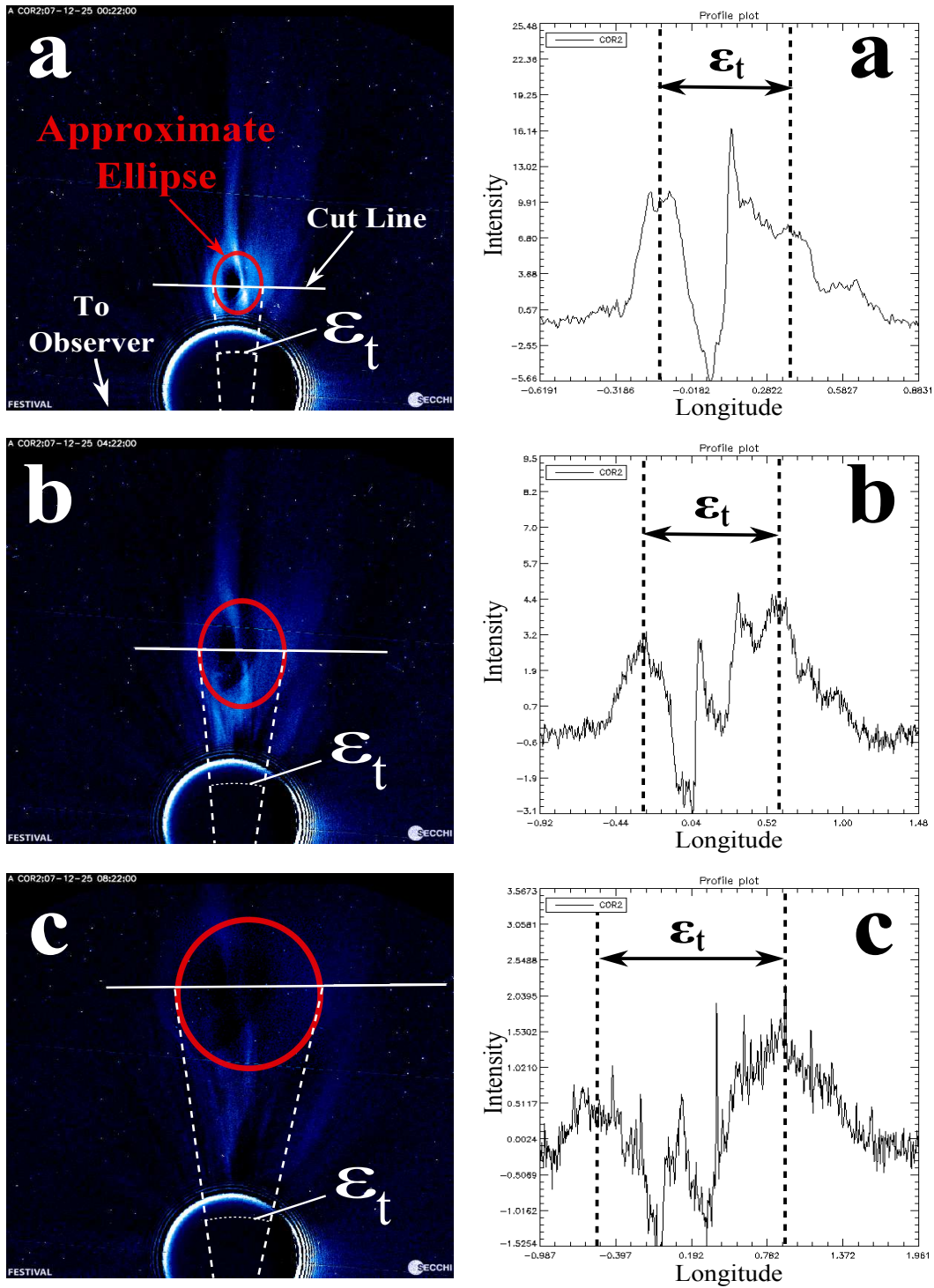}
			\end{center}
			\caption{Rotated images of a CME from STEREO A (left) alongside their accompanying intensity profiles (right) taken at (a) 0:22, (b) 4:22, and (c) 8:22 (HH:MM) on 25 December 2007. The intensity profiles are plots of intensity as a function of longitude (in the horizontal direction) along the white cut line shown in the images on the left. An approximate outline of the ellipse fit to each image and the elongation of the transverse width, $\epsilon_t$, (the angle between the lines connecting the ``sides" of the CME to the observer) are overlaid on the images. The elongation of the transverse width is also indicated by the bounded arrow in the intensity profiles, with the bounds set at the center of each large density structure. The elongation measurements are in degrees and the intensity measurements are in units of solar disk intensity (linear scale).}
			\label{fig:Fig3}
		\end{minipage}
\end{figure}

We obtained level 0.5 images (the decompressed data in FITS format) from the
 STEREO and SOHO servers. FESTIVAL, an IDL-based browser designed for
 manipulating solar imaging data \cite{Auchere08}, applied a daily
 background subtraction and calibration to each image in order to generate the
 images we used, shown in Figures~\ref{fig:Fig2a} and~\ref{fig:Fig3}.
\par
To fit the observed elliptical cross section, four measurements were acquired
 from each image. These measurements were $\epsilon_{LE}$ and $\epsilon_{IC}$,
 the elongations of the leading edge and inner core of the CME (the feature-observer-Sun angle), 
$\theta$, the position angle (the angle between ecliptic
 north and the Sun-CME center line in the image), and the transverse half-width,
 $w_{t}$. Figure~\ref{fig:Fig2a} displays the first three measurements on an image from STEREO A taken
 on 25 December 2007. The fourth, the transverse half-width, is shown in Figure~\ref{fig:Fig3}.
 The distance from the Sun to the observing satellite, $d_0$, and the
 propagation angle, $\phi$, (the observer-Sun-CME center angle) which will be
 discussed later in the paper, are also shown. Figure~\ref{fig:Fig2b} is a plot generated by
 FESTIVAL of the intensity along the cut through the image on the diagonal white
 line in Figure~\ref{fig:Fig2a} (termed ``cut line"), and will herein be referred to as an
 ``intensity profile". The data spike near the longitude of -2.14 degrees is 
due to the cut line crossing over a bright star in the image. Because of the
 high detail of the SECCHI imagers, the intensity profile varied greatly with
 the location of the cut line. To reduce this variation, we modified the
 FESTIVAL code to plot the average intensity within $\pm$10 pixels of the cut 
line. This modification produced more reliable results that were less dependent
 upon where the cut line was drawn.

%

\par
The elongations measured from each intensity profile were converted into distance
 measurements using the ``fixed-$\phi$" approximation \cite{Kahler07}:
\begin{equation} \label{eq:Eq1}
r = d_0 \frac{\sin(\epsilon)}{\sin{(\epsilon+\phi)}},
\end{equation}
where is the elongation of a particular feature. Equation (1) does not account for
 the effects of the Thomson sphere or the possibility of a wide, 3D structure in
 a CME \cite{Lugaz09,Wood11}. Because our measurements were confined to
 elongations less than $16\,^{\circ}$ and radial distances less than .3 AU, the errors due
 to disregarding the Thomson sphere and structure of the CME are negligible
 \cite{Vourlidas06,Howard09a}. Other approaches, such as the ``harmonic mean",
 have been used by other groups to translate elongations into radial distances
 \cite{Lugaz09}. These techniques have proven to be better suited for specific
 types of CMEs, namely broad, bright CME fronts \cite{Wood10}. To keep our approach
 consistent, we used the ``fixed-$\phi$" method in our model for each event.
\par
The intensity profiles in FESTIVAL measured intensity as a function of longitude
 (only in the horizontal direction in the image). Therefore, to properly measure
 the transverse width we had to rotate each image until the transverse width was
 completely horizontal. We then created an intensity profile across the entire 
width of the CME to measure its transverse size. 

Figure~\ref{fig:Fig3} displays three
 examples of the rotated CME images from STEREO A (left) alongside their
 accompanying intensity profiles (right) taken at (a) 00:22, (b) 04:22, and (c)
 08:22 (HH:MM UT) on 25 December 2007. An approximate outline of the ellipse we
 fit to each image and the elongation of the transverse width are overlaid on
 the images. The elongation of the transverse width (the angle between the two 
lines connecting the ``sides" of the CME to the observing satellite) is also
 indicated by the bounded arrow in the intensity profiles. The intensity profile
 in Figure~\ref{fig:Fig3}a shows how the determination of the $\epsilon_t$ can be ambiguous when using
 a single image. To determine the appropriate features to use in measuring 
$\epsilon_t$, a sequence of images was examined and features that remained
 bright over a large range of elongations were selected. The bounds on the
 transverse elongation (the dotted lines in the right panels of Figure~\ref{fig:Fig3}) were
 set at the middle of the large density enhancements in the intensity profiles.
 The transverse elongation was converted to a distance measurement using
\begin{equation} \label{eq:Eq2}
w_t = d_0 \frac{\sin(\phi)}{\sin{(\phi+\epsilon_c)}}\tan{\left(\frac{\epsilon_t}{2}\right)},
\end{equation}
where $\epsilon_c$ is the elongation of the CME center and $w_t$ is the
 transverse half-width of the CME. Equation (2) is derived in the appendix.
\par
FESTIVAL was used to calculate the position angle, $\theta$, of the CME 
in each frame by measuring the location of the CME center with respect 
to the Sun. However, the position angle was slightly different from the 
latitudinal angle, $\theta^\prime$, (the angle between ecliptic north and
 the Sun-CME center line in 3D space). Projection effects had to be 
considered to correctly determine the trajectory of the CME. Equation (3)
 is derived in the appendix and gives the latitudinal angle by
\begin{equation} \label{eq:Eq3}
\theta^\prime = \arcsin{\left(\sqrt{\sin^2(\theta) \sin^2(\phi) + \cos^2(\phi)}\right)}.
\end{equation}
\par
To determine the propagation angle, $\phi$, of each CME, we compared 
near-simultaneous images, taken less than ten minutes apart, from the
 two observing satellites. Assuming that the features measured in both
 sets of images were the same, the radial distances associated with both
 satellite elongation data sets should be equal, even though the respective
 elongation values were different. We compared the inner core data sets 
from both observing satellites (between 15-30 data points for the events in
 our study) and determined the value for $\phi$ which resulted in the smallest 
sum of the absolute errors between the radial distances in both data sets. We 
then compared the leading edge data sets from both observing satellites in a 
similar manner to determine a second value for $\phi$. We then used the
 average of both $\phi$ values to obtain a final propagation angle for each 
event. In other words, the selected value for $\phi$ was the one that resulted 
in the least error between the leading edge and inner core data sets for both 
observing satellites.
\par
Our methodology can be summarized as follows: 
\begin{enumerate}
\item[1)] Measure the variables which characterize the propagation and expansion of
 the CME ($w_r$, $w_t$, $r_c$, $\theta$, and $\phi$) using the density structures
 seen in the imager data.
\item[2)] Generate a mathematical 3D ellipsoid surface (as a function of time) using 
the measured variables in step 1 (see Equation 5).
\item[3)] Calculate the predicted arrival times in the \emph{in situ} data using the
 time-dependent mathematical ellipsoid.
\end{enumerate}

\section{Using Imager Data to Generate Predictions} 
      \label{S-3}
\par
Using the methodology described above, we now examine the propagation of three 
events. The center of each CME was calculated as the mean of the leading edge 
and inner core radial positions. A line was fit to the data set for the center 
of each event to predict its location beyond the distance of the image data. 
For event 2, a straight line was fit to the data after about 19:00 on April 26th,
 because its inner core experienced a brief acceleration at that point in time. 
A straight line was fit to the entire data set from event 3. Event 1 exhibited 
a slow initial velocity ($\le$100 km/s) and gradually ``ballooned" off the surface of
 the Sun. This initial behavior typically leads to extended, slow acceleration
 \cite{Sheeley99} and required a different fitting technique because it did not
 propagate with constant speed. Although its acceleration decreased within the 
radial distance over which images were obtained, it was still significant enough 
to strongly affect the estimated arrival time; a linear fit of the last 15 data 
points produced an arrival time that was nearly 24 hours late when compared with
 the \emph{in situ} data. Following the example of \inlinecite{Sheeley99}, we fit 
\begin{equation} \label{eq:Eq4}
R(t) = r_0 + 2r_a \ln{\left[\cosh{\left(\frac{v_a (t+\Delta t_0)}{2r_a}\right)}\right]}
\end{equation}
to the data series for event 1, in which $R(t)$ is the radial distance of the 
feature as a function of time, $r_0$ is radial position at $t = -\Delta t_0$,
$v_a$ is the asymptotic velocity, and $r_a$ is the e-folding distance at which
 the asymptotic velocity is approached. Instead of measuring these values 
independently, we optimized $r_a$, $r_0$, and $v_a$ in order to achieve a quality fit.
\par
Table~\ref{tab:Tab1} displays the variables which characterize the trajectory of the CME. The
 columns, from left to right, list the propagation angle (the observer-Sun-CME
 center angle), the latitudinal angle (between ecliptic north and the Sun-CME center 
line), the asymptotic velocity, and the e-folding distance at which the asymptotic
 velocity is approached. For events 2 and 3, the asymptotic velocity was taken as 
the velocity from the linear fit of the data. {\em Note: The latitudinal angle 
($\theta^\prime$) of Event 2 changed as it propagated away from the Sun (see Figure~\ref{fig:Fig6}).
 The value listed in Table~\ref{tab:Tab1} is the latitudinal angle at the time it crossed} 1 AU.
\par
\begin{figure}[hp]
	\begin{center}
		\begin{minipage}[t]{0.68\textwidth}
			\begin{center}
				\includegraphics[width=\textwidth]{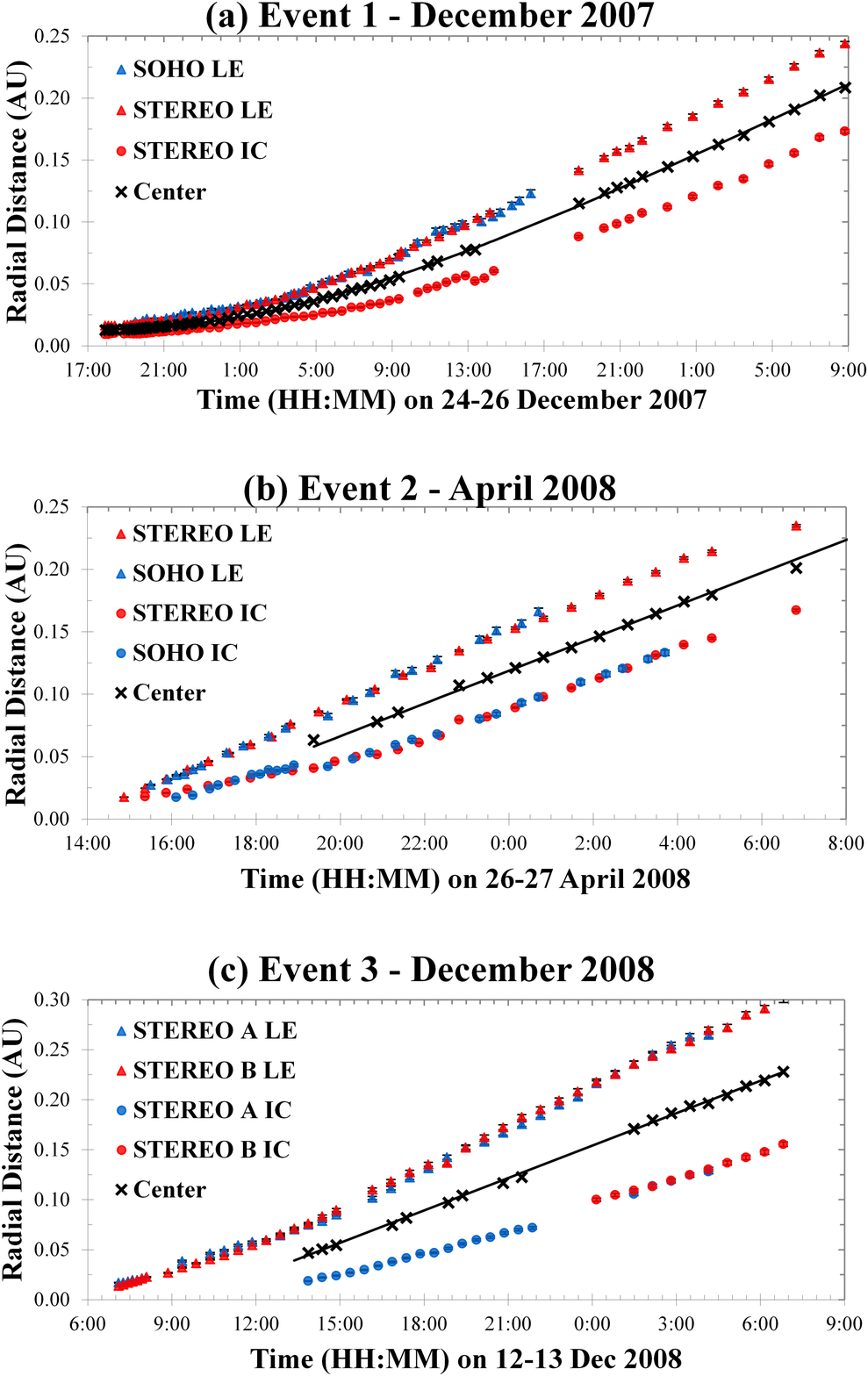}
			\end{center}
			\caption{A height-time diagram (plot of radial distance vs. time) for the leading edge and inner core of the CME for events 1 (a), 2 (b), and 3 (c). The CME center and the respective fit line for each event are also plotted. Where not visible, the error bars on each data point are approximately the size of the marker.}
			\label{fig:Fig4}
		\end{minipage}
	\end{center}
\end{figure}

Figure~\ref{fig:Fig4} displays the radial distance of the leading edge (LE) and inner core (IC)
 as a function of time for each event, also known as a height-time diagram 
\cite{Sheeley99}. Additionally, the center of each event, along with the respective 
fit line, is plotted. The data obtained from the SOHO and STEREO images, displayed in 
different colors, demonstrates the good agreement between both data sets.
\begin{table}[htp]
	\caption{ The variables associated with the kinematics of the three CMEs in our study. $v_a$ is 
the asymptotic velocity of the CME center, $r_a$ is the e-folding distance at which the asymptotic
 velocity is approached, $\phi$ is the observer-Sun-CME center angle (from STEREO A), and 
$\theta^\prime$ is the angle between ecliptic north and the Sun-CME center line.}
	\label{tab:Tab1}
	\begin{tabular*}{5in}{@{\extracolsep\fill}ccccc}
		\hline
		Event & $\phi$ (degrees) & $\theta^\prime$ (degrees) & $v_a$ (km/s) & $r_a$ (AU)\\
		\hline
		1 &  58.15  & 93.64 & 308.85 & .068\\
		2 &  61.70  & 85.10 & 584.42 & n/a\\
		3 &  41.40  & 84.94 & 578.89 & n/a\\
		\hline
	\end{tabular*}
\vspace{-6mm}
\end{table}
\newpage
\begin{figure}[ht]
	\begin{center}
		\begin{minipage}[t]{0.68\textwidth}
			\begin{center}
				\includegraphics[width=\textwidth]{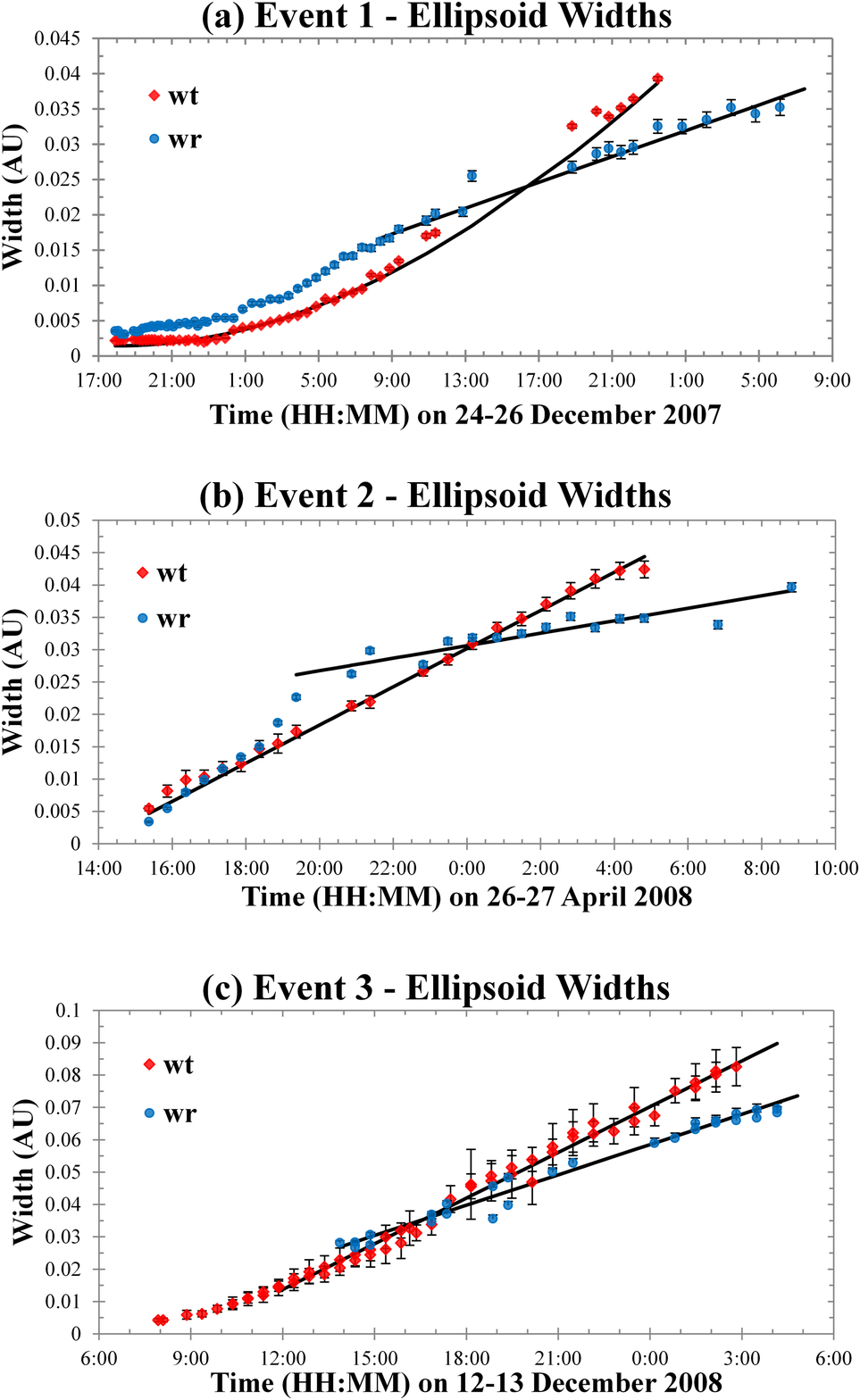}
			\end{center}
			\caption{A plot of the radial CME half-width along the Sun-CME center line ($w_r$) and the transverse half-width ($w_t$) of the ellipse fit to each image for events 1 (a), 2 (b), and 3(c). The trend lines used to predict the growth of both half-widths are also shown. Where not visible, the error bars on each data point are approximately the size of the marker.}
			\label{fig:Fig5}
		\end{minipage}
	\end{center}
\end{figure}
\par
The radial half-width of the ellipsoid, $w_r$, was calculated by taking half the distance
 between the leading edge and inner core. Figure~\ref{fig:Fig5} displays the growth of the radial 
half-width and the transverse half-width, $w_t$, for each event, as well as respective 
fit used on each data set. 
\par
It was necessary to study the position angle ($\theta$) as a function of time for 
each CME. As events 1 and 3 propagated outwards, $\theta$ remained relatively constant
 with small fluctuations about the average, attributable to the user-related error
 introduced by our technique. In contrast, the position angle for event 2 was
 initially $\approx 70\,^{\circ}$ and then increased significantly. To describe this
 trend, we performed a least squares fit with an arctangent function, as shown in
 Figure~\ref{fig:Fig6}. This trend-line fit the data quite well and indicated that event 2 did
 not propagate along a purely radial trajectory.
\begin{figure}
	\begin{center}
		\begin{minipage}[b]{0.85\textwidth}
			\begin{center}
				\includegraphics[width=\textwidth]{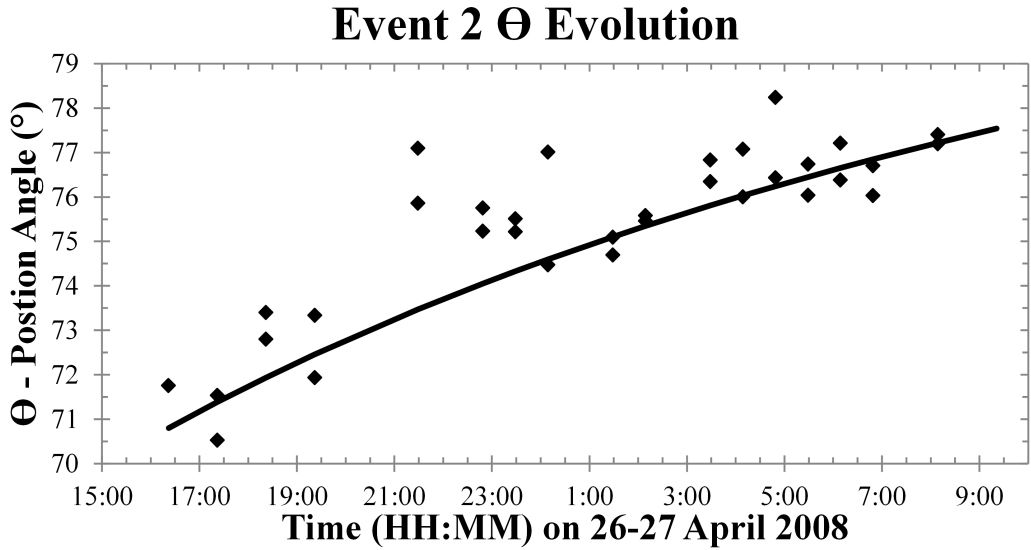}
			\end{center}
			\caption{The evolution of the position angle ($\theta$) for event 2 as it propagated away from the Sun. The accompanying trend line was generated by a least squares fit of an arctangent function. The effectiveness of this fit inidicates that event 2 did not propagation in a purely radial direction.}
			\label{fig:Fig6}
		\end{minipage}
	\end{center}
\end{figure}
\par
To determine the arrival time for each event in our study, we defined a modified 
radial, tangential, normal (RTN) coordinate system, with the R-axis pointing from 
the Sun center to the CME center, the N-axis being perpendicular to the R-axis and 
in the same plane as the ecliptic north, and the T-axis completing the orthogonal 
coordinate system (Figure~\ref{fig:Fig2a}). In other words, our coordinate system was defined using
 the propagation angle ($\phi$) and the latitudinal angle ($\theta^\prime$) and was
 designed such that each width used to define our ellipsoid was directed along an
 axis. Then, using the time-dependent location of the CME center ($r_c$) and its 
radial, transverse, and normal half-widths ($w_r$, $w_t$, and $w_n$), we 
mathematically modeled the 3D surface of the ellipsoid. This surface is shown in 
Figure~\ref{fig:Fig1} and is defined as
\begin{equation} \label{eq:Eq5}
\left(\frac{r-r_c}{w_r(t)}\right)^2 + \left(\frac{t}{w_t(t)}\right)^2 + \left(\frac{n}{w_n(t)}\right)^2 = 1,
\end{equation}
in which ($r$, $t$, $n$)  are the coordinates of the surface. As previously stated,
 we assumed that the transverse half-width was equal to the normal half-width 
($w_t$ = $w_n$). This assumption is examined in detail in the conclusion. Using the
 position of the transited satellite and the equation describing the time-dependent 
surface of the ellipsoid in the new RTN coordinate system, we calculated when the 
leading and trailing edge of the ellipsoid would traverse the satellite. These two
 arrival times were compared to the \emph{in situ} data from the transited satellite.

\section{Comparisons to \emph{In Situ} Data}
      \label{S-4}
\par
\begin{figure}[ph]
	\centering
		\begin{minipage}{0.65\textwidth}
			\begin{center}
				\includegraphics[width=\textwidth]{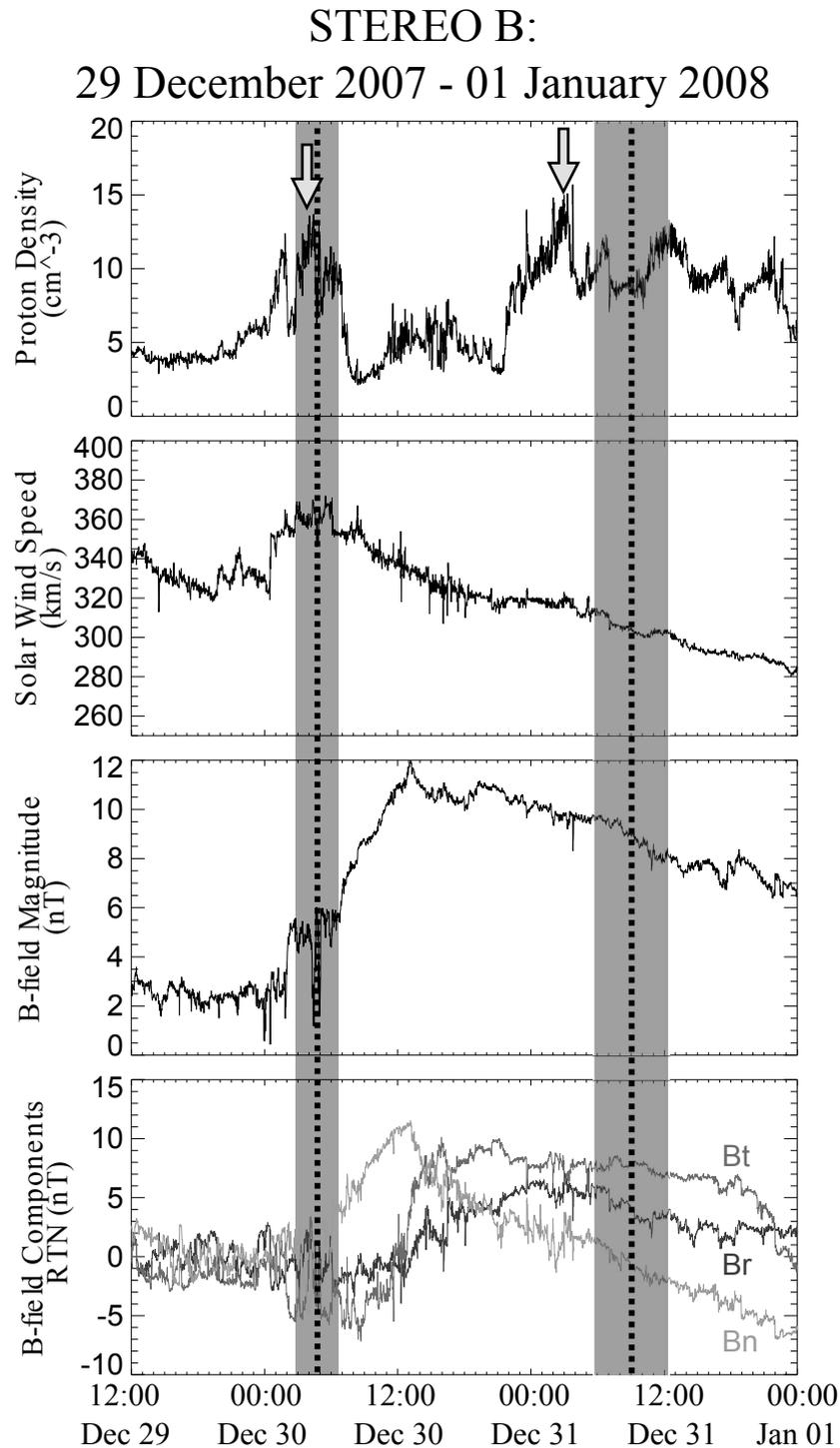}
			\end{center}
			\caption{Event 1 in the in situ data from 30 December 2007 to 01 January 2008, as observed by STEREO B when it was about $23\,^{\circ}$ off the Sun-Earth line. The panels, from top to bottom, display the proton density, solar wind speed, magnitude of the magnetic field and the components of the magnetic field in an RTN coordinate system. The predicted arrival times from our ellipsoid model are marked by the dotted black lines and bounded by $\pm1\sigma$ (grey). The arrows overlaid on the density plot mark the enhancements we compared to our model, namely the first enhancement and the one directly following the low-density cavity.}
			\label{fig:Fig7}
		\end{minipage}
  \end{figure}
\begin{figure}[ph]
	\centering
		\begin{minipage}[b]{0.65\textwidth}
			\begin{center}
				\includegraphics[width=\textwidth]{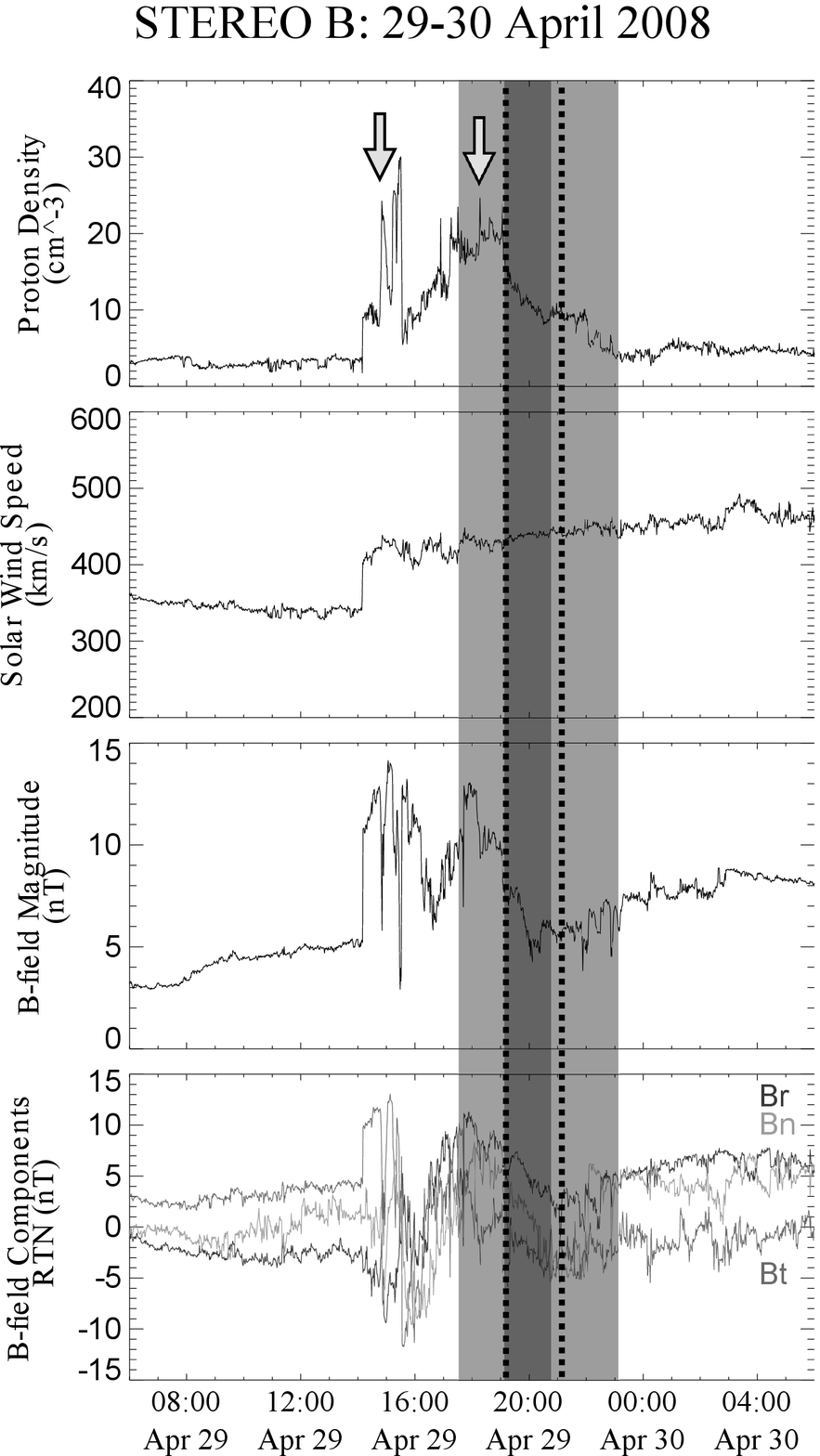}
			\end{center}
			\caption{Event 2 in the in situ data from 29 April 2008 to 30 April 2008, as observed by STEREO B when it was about $24\,^{\circ}$ off the Sun-Earth line. The panels, from top to bottom, display the proton density, solar wind speed, magnitude of the magnetic field and the components of the magnetic field in an RTN coordinate system. The predicted arrival times from our ellipsoid model are marked by the dotted black lines and bounded by $\pm1\sigma$ grey/black boxed). The $\pm1\sigma$ bounds overlap and are shaded darker in the overlapping region. The arrows overlaid on the density plot mark the enhancements we compared to our model, namely the first enhancement and the one directly following the low-density cavity.}
			\label{fig:Fig8}
		\end{minipage}
\end{figure}
\begin{figure}[ph]
	\centering
		\begin{minipage}[b]{0.65\textwidth}
			\begin{center}
				\includegraphics[width=\textwidth]{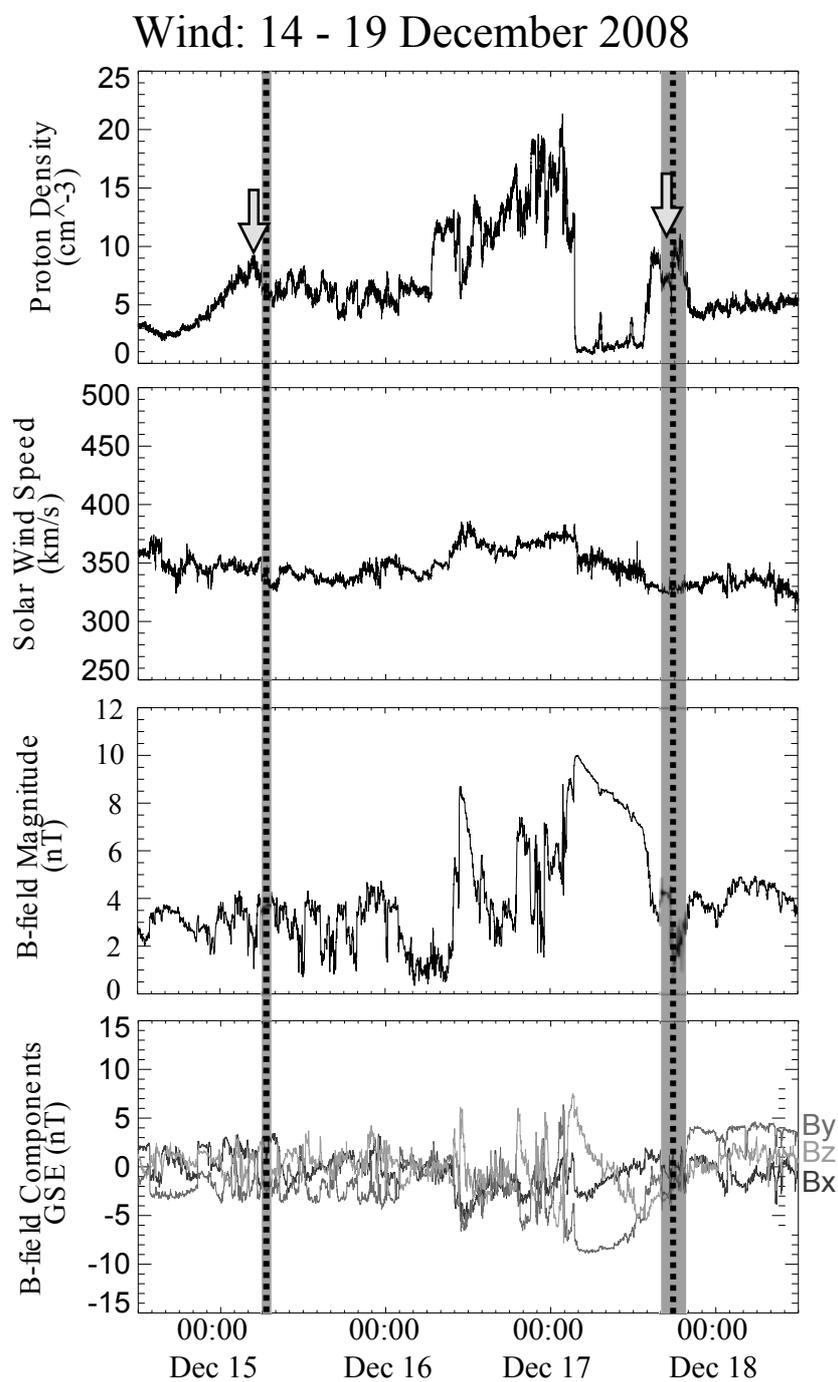}
			\end{center}
			\caption{Event 3 in the in situ data from 14 December 2008 to 19 December 2008, as observed by Wind when it was roughly 1.5 million kilometers upstream of the Earth. The panels, from top to bottom, display the proton density, solar wind speed, magnitude of the magnetic field and the components of the magnetic field in an RTN coordinate system. The predicted arrival times from our ellipsoid model are marked by the dotted black lines and bounded by $\pm1\sigma$ (grey). The arrows overlaid on the density plot mark the enhancements we compared to our model, namely the first enhancement and the one directly following the low-density cavity. Though this is preliminary browse data, it agrees well with similar data from the Advanced Composition Explorer (ACE).}
			\label{fig:Fig9}
		\end{minipage}
\end{figure}

Coronal mass ejections are characterized by a variety of \emph{in situ} signatures 
including changes in the density, the magnitude and direction of the magnetic field,
  and the velocity of the solar wind \cite[and references therein]{Jian06}. Figures~\ref{fig:Fig7},~\ref{fig:Fig8}, and~\ref{fig:Fig9} display the \emph{in situ} data for each event over-plotted with our 
predicted arrival times. The arrows in each density plot identify the density 
enhancements we compared to our ellipsoid model. We chose to compare our model to the
 first density enhancement and the enhancement directly following the low-density 
cavity. The first density enhancement was selected because the ellipsoid model was 
fit to the leading edge, which could reasonably be expected to cross the \emph{in situ} 
observer first. The enhancement following the low-density cavity was chosen because 
the ellipsoid was also fit through the inner core, which directly followed the dark,
 low-density cavity in the images. Figure~\ref{fig:Fig7} shows that multiple density enhancements 
occurred after the feature to which we compared our predictions. This highlights the 
fact that the predictions from the ellipsoid model do not accurately describe density
 enhancements sunward of the inner core, as stated in the introduction of this paper.
 
 \par
Table~\ref{tab:Tab2} displays the comparison between the predictions generated by our model and 
the density enhancements in the \emph{in situ} data. The columns, from left to right, 
identify the event number and transited satellite, the predicted time of the 
beginning/end of the model transit, the time of the density enhancement in the 
\emph{in situ} data, and difference between the predicted and observed arrival time
 (in hours and standard deviations). The uncertainty for the \emph{in situ} time was
 calculated as half the temporal width of the density enhancement.
\par

\begin{table}[ht]
\caption{The predicted beginning and end of the ellipsoid’s transit compared to the first density enhancement and the one directly following the low-density cavity in the \emph{in situ} data, respectively. The transited satellite and event number are recorded in the left-hand column for three coronal mass ejections.}
\label{tab:Tab2}
\begin{tabular}{cllcc}
  \hline
  Event/satellite & Predicted Arrival Time & \emph{In Situ} Arrival Time & Hours Away & $\sigma$ Away\\
  \hline
  1 & 12/30/07 4:56 $\pm$ 2:07 & 12/30/07 4:00 $\pm$ 3:00 & 0.95 & 0.26\\
  ST-B & 12/31/07 9:09 $\pm$ 3:17 & 12/31/07 3:00 $\pm$ 0:45 & 6.16 & 1.83\\
  \hline
  2 & 4/29/08 19:12 $\pm$ 1:38 & 4/29/08 14:45 $\pm$ 0:45 & 4.46 & 2.47\\
  ST-B & 4/29/08 21:08 $\pm$ 2:05 & 4/29/08 18:15 $\pm$ 1:00 & 2.90 & 1.25\\
  \hline
  3 & 12/15/08 6:45 $\pm$ 0:20 & 12/15/08 4:30 $\pm$ 3:00 & 2.26 &0.75\\
  Wind & 12/17/08 17:41 $\pm$ 1:08 & 12/17/08 17:15 $\pm$ 3:00 & 0.44 & 0.14\\
  \hline
\end{tabular}
\vspace{-6mm}
\end{table}

\par
To further test our model, we compared the temporal separation of the density peaks
 tabulated in Table~\ref{tab:Tab2}. Table~\ref{tab:Tab3} displays the predicted temporal separation to that 
seen in the \emph{in situ} data. The far right column displays the difference between
 the predicted duration from the model and the duration in the \emph{in situ} data. 
The average accuracy of these predictions was 2.6 hours.

\begin{table}[ht]
\caption{The predicted temporal duration of each CME compared with the temporal separation seen in the \emph{in situ} data, defined as the time between the first density enhancement and the one following the low-density cavity.}
	\label{tab:Tab3}
	\begin{tabular}{cccc}
	\hline
		Event & Ellipsoid Width (hours) & \emph{In situ} Width (hours) & Error\\
		\hline
		1 & 28.22$\pm$3.91 & 23.00 & 5.22\\
		2 & 1.94$\pm$2.67 & 3.50 & -1.56\\
		3 & 58.93$\pm$1.19 & 60.75 & -1.82\\
			\hline
	\end{tabular}

\vspace{-6mm}
\end{table}

\section{Summary and Conlusions}
    \label{S-5}
\par
Although tracking and predicting the propagation of coronal mass ejections using 
images is nothing new, our study introduces two tools that could prove useful in 
the future study of CMEs. First, we demonstrate the use of intensity profiles in 
FESTIVAL (Figures~\ref{fig:Fig2b} and~\ref{fig:Fig3}) as an alternative way to construct height-time diagrams
 and study the transverse growth of CMEs. Previous methods using image profiles 
have been constrained to constant position angles \cite{Sheeley08}. The intensity
 profiles implemented in FESTIVAL are not constrained in this manner and can 
therefore track CMEs with varying position angles (such as seen in event 2). 
Additionally, intensity profiles provide a method for determining the uncertainty
on each measurement (see Figure~\ref{fig:Fig2b}), instead of assuming a certain percent error 
in each measurement based on the imager used \cite{Wood09,Lugaz10,Wood10}.
\par
Second, we introduce our ellipsoid model and test it on three separate CMEs.
 The results suggest that the ellipsoid model may prove useful in tracking 
and predicting the expansion and trajectory of CMEs. Our predictions were an 
average of 2.9 hours away from the arrival of particular density enhancements 
in the \emph{in situ} data for the three events. The average of the uncertainty
 in our predicted arrival times was 1.9 hours and compares well with previous 
work \cite[see their Table 2]{Davis09}. \inlinecite{Howard09b} and 
\inlinecite{Wood11} used synthetic images obtained from their 3D CME density
 models to track the kinematics of CMEs and, when compared to the \emph{in situ} 
data, achieved ideal accuracies of 3.5 hours and less than an hour, respectively.
\inlinecite{Lugaz10} used a model that accounted for the 3D structure of CMEs 
and obtained an accuracy of 11 hours. Using the synthetic image technique, 
\inlinecite{Wood09} predicted the \emph{in situ} signatures of the April 2008 
CME (our event 2) with accuracies ranging from 1.5 to 10 hours. Though our 
ellipsoid model is as accurate as these previous studies, this not an entirely
 appropriate comparison. The aforementioned models incorporated measurements 
from the Sun to 1 AU, except \inlinecite{Lugaz10}, who used data out to 
elongations of about $45\,^{\circ}$. The fact that our predictions were obtained
 using data closer to the Sun (measurements below .3 AU and elongations less 
than $16\,^{\circ}$) and attained a similar accuracy is a promising result. Our
 ellipsoid model demonstrates the possible potential of being able to accurately
 predict CME effects when the CME is still days away from reaching 1 AU.
\par
Some of the errors given in Tables~\ref{tab:Tab2} and~\ref{tab:Tab3} can be attributed to our neglect of
 the effects of the Thomson sphere and the 3D structure of the CME. Another 
plausible source of error originates from using an ellipse of revolution 
around the radial axis, making our model azimuthally symmetric. This meant
 that the third width of our ellipsoid model (approximately along the imager’s
 line-of-sight) was set equal to the transverse width. The use of an 
azimuthally symmetric structure (an ellipse of revolution) is not intended to
 represent an accurate physical assumption, but rather arises from the 
limitations of the two vantage points used in our study. To circumvent this, 
one could examine images from the transited satellite to measure this third width.
 Additionally, the fact that all of the predictions of our model were later than 
the actual \emph{in situ} times may prove useful in understanding the physics of
 CME propagation between .3 and 1 AU.
\par
The study of three CMEs also shed light on a few assumptions made in some 
current approaches. Past studies have often assumed that CMEs travel with c
onstant velocity throughout the HI-1 field of view \cite{Davis09,Savani09,Liu10},
 which begins at an elongation of $4\,^{\circ}$, or $\approx$.07 AU for the events 
in our study. Event 1 exhibited significant acceleration beyond an elongation of
 $14\,^{\circ}$, demonstrating that this assumption does not hold true for all CMEs.
\inlinecite{Savani09}used a circular cross section to model the flux rope geometry.
 Figure~\ref{fig:Fig5} shows that the circular cross section assumption is roughly true close to 
the Sun, but as the CME propagates away from the Sun, the two radii observed in 
the images become significantly different. Additionally, an examination of the 
transverse, or normal, growth of events 1 and 2 (Figure~\ref{fig:Fig5}) showed that 
the assumption of self-similar expansion does not hold true for all CMEs. Though
 many CMEs certainly exhibit self-similar expansion, our results indicate that this
 assumption can be made only after a careful examination of the data.
\par
CMEs are a diverse group of structures in the solar wind and require a flexible tool
 to adequately track and predict their propagation. Their radial velocity and 
transverse growth can accelerate beyond .25 AU (event 1), their expansion rates 
can drastically change (event 2) as far out as .08 AU, and their direction of
 propagation can change over time (event 2). Our ellipsoid model provides a method 
to incorporate this variety using a small number of parameters. After examining three
 CMEs using the ellipsoid model, we find it capable of accurately predicting the
 occurrence of density enhancements and the duration of an event in the density data.
 Though this is not a statistical study, the results show some promise that the 
forward structure of a CME, from the inner core to the leading edge, can be 
effectively modeled as an ellipsoid.

\section{Acknowledgements}
    \label{S-Ack}
\par
STEREO/HI was developed by a consortium comprising RAL, the University of 
Birmingham (UK), CSL (Belgium) and NRL (USA). SECCHI, led by NRL, involves
 additional collaborators from LMSAL, GSFC (USA), MPI (Germany), IOTA and 
IAS (France). We thank the PIs of the STEREO, SOHO and Wind instruments 
utilized herein: R. A. Howard, G. Brueckner, J. Luhmann, M. Acuna, A. Galvin,
 R. Lin and R. Lepping. We would like to thank Elie Soubri$\grave{e}$ and
 Fr$\acute{e}$d$\acute{e}$ric Auch$\grave{e}$re of the Institut 
d’Astrophysique Spatiale (IAS) for assisting with the modifications 
of FESTIVAL’s code. We thank Dr. Aaron Breneman for reviewing the paper, 
Dr. Dan Cronin-Hennessy for his insightful conversations concerning the 
propagation of uncertainties, and Ying Liu for feedback regarding our methods.
\par 
This research was supported through a grant from the University of Minnesota 
Undergraduate Research Program and by NASA grant NNX09AG82G.

\renewcommand{\theequation}{A\arabic{equation}}\setcounter{equation}{0}
\section*{Appendix: Deriving Equations (2) and (3)}
\par
\begin{figure}[ht]
	\begin{center}
		\begin{minipage}[b]{0.65\textwidth}
			\begin{center}
				\includegraphics[width=\textwidth]{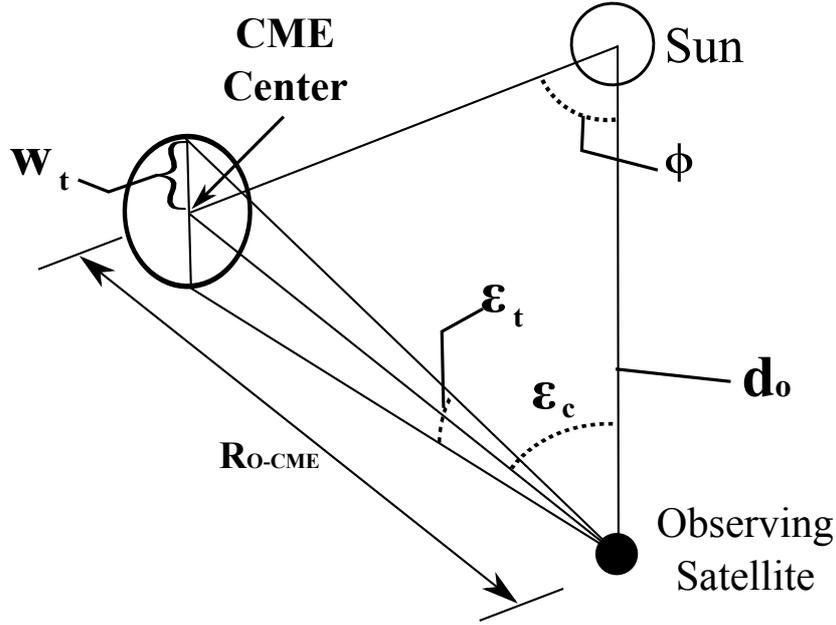}
			\end{center}
			\caption{The geometry of an arbitrary CME propagation with the ellipsoid cross section shown. The method for converting the elongation width of the CME ($\epsilon_t$) to the distance half-width of the CME ($w_t$) is displayed. Also shown is the elongation of the CME center ($\epsilon_c$), the distance from the observer to the Sun ($d_0$), the distance from the observer to the CME center ($R_{O-CME}$), and the propagation angle ($\phi$).}
			\label{fig:FigA1}
		\end{minipage}
	\end{center}
\end{figure}

 
Equation 2 is derived from the geometry presented in Figure~\ref{fig:FigA1}, in which $w_t$
 is half the transverse distance width of the CME, $\epsilon_t$ is the 
elongation of the transverse width of the CME, $\epsilon_c$ is the elongation
 of the CME center, $d_0$ is the distance from the observer to the Sun, $R_{O-CME}$ 
is the distance from the observer to the CME center, and $\phi$ is the propagation 
angle (the observer-Sun-CME center angle). 
Using the sine law on the triangle connecting the Sun, observer, and CME center yields
\begin{equation}
\sin{(\phi)} d_0 = \sin{(180-\phi-\epsilon_c)} R_{O-CME}.
\end{equation}
Applying the right triangle rule to the triangle connecting the CME center, observer, 
and CME side results in
\begin{equation}
\tan{\left(\frac{\epsilon_t}{2}\right)} = \frac{w_t}{R_{O-CME}}.
\end{equation}
Combining (A1) and (A2) produces Equation (2)
\begin{equation}
w_t = R_0 \frac{\sin{(\phi)}}{\sin{(\phi+\epsilon_c)}} \tan{\left(\frac{\epsilon_t}{2}\right)}.
\end{equation}
\par
\begin{figure}[ht]
	\begin{center}
		\begin{minipage}[b]{0.65\textwidth}
			\begin{center}
				\includegraphics[width=\textwidth]{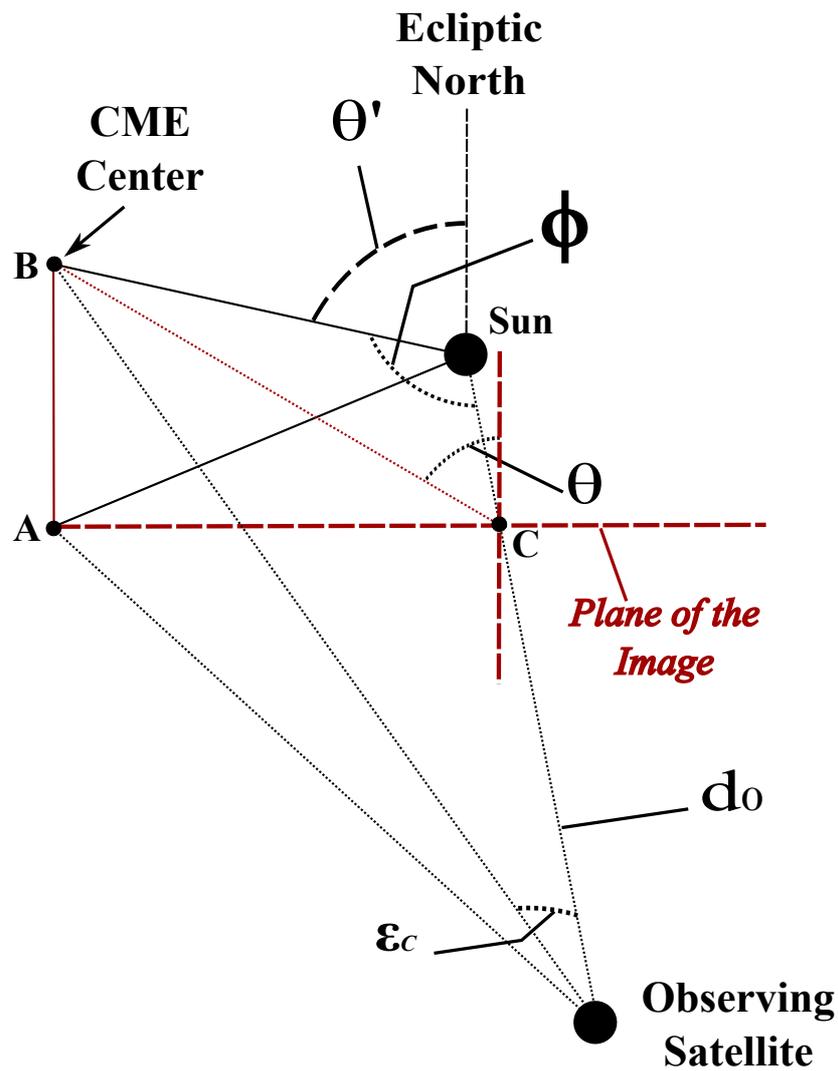}
			\end{center}
			\caption{The geometry of an arbitrary CME propagation. The relationship between the position angle ($\theta$) as viewed from the image and the latitudinal angle ($\theta^\prime$) in 3D space is displayed. Also shown is the elongation of the CME center ($\epsilon_c$), the propagation angle ($\phi$), and the plane onto which all features are projected to form the image seen by the observing satellite (``Plane of the Image").}
			\label{fig:FigA2}
		\end{minipage}
	\end{center}
\end{figure}

Equation 3 is derived from the geometry presented in Figure~\ref{fig:FigA2}, in which the
 ``Plane of the Image" is an arbitrary plane onto which all features are projected
 to form the image seen by the observing satellite. $\theta^\prime$ is the 
latitudinal angle, $\theta$ is the position angle seen in the image, $\epsilon_c$
 is the elongation of the CME center, $\phi$ is the propagation angle, and $d_0$
is the distance from the observer to the Sun. 
Applying the right triangle rule to the triangle connecting A, B, and the Sun (S)
 yields (``AS" is the length of the line connecting A and the Sun)
\begin{equation}
\sin{(\theta^\prime)} = \frac{AS}{BS}.
\end{equation}
Defining $\phi^\prime$ as the A-Sun-C angle, we apply the right triangle rule
 to ABC and to BCS to get
\begin{equation}
\sin(\theta) = \frac{AC}{BC} = \frac{AS\sin{(\phi^\prime)}}{BS\sin(\phi)}.
\end{equation}
$\phi^\prime$ can be expressed in terms of $\phi$ and $\theta^\prime$:
\begin{equation}
\phi^\prime = \arccos \left(\frac{\cos(\phi)}{\sin{(\theta^\prime)}}\right).
\end{equation}
We use the right triangle rule to define from (A6) and then combine it 
with (A4) and (A5) to get
\begin{equation}
\sin{(\theta^\prime)} = \frac{\sin(\theta) \sin(\phi)}{\sqrt{1 - \frac{\cos^2(\phi)}{\sin^2{(\theta^\prime)}}}}.
\end{equation}
Multiplying both sides by the quantity in the square-root and then squaring 
both sides yields
\begin{equation}
\sin^2(\theta^\prime) - \cos^2(\phi) = \sin^2(\theta) \sin^2(\phi).
\end{equation}
Solving for  we obtain the final expression, Equation (3): 
\begin{equation}
\theta^\prime = \arcsin{\left(\sqrt{\sin^2(\theta)\sin^2(\phi) + \cos^2(\phi)}\right)}.
\end{equation}

\newpage
\bibliographystyle{spr-mp-sola}
  
\bibliography{Sam_EndNote_References}  

\IfFileExists{\jobname.bbl}{} {\typeout{}
\typeout{****************************************************}
\typeout{****************************************************}
\typeout{** Please run "bibtex \jobname" to obtain} \typeout{**
the bibliography and then re-run LaTeX} \typeout{** twice to fix
the references !}
\typeout{****************************************************}
\typeout{****************************************************}
\typeout{}}

\end{article} 
\end{document}